\documentclass[aps,amsfonts,amsmath,tightenlines,showpacs,nofootinbib]{revtex4}
\usepackage{epsf}

\def\be{\begin{equation}}
\def\ee{\end{equation}}
\def\beq{\begin{equation}}
\def\eeq{\end{equation}}
\def\bea{\begin{eqnarray}}
\def\eea{\end{eqnarray}}
\def\bml{\begin{subequations}}
\def\blea{\bml\begin{eqnarray}}
\def\elea{\end{eqnarray}\end{subequations}}

\begin{document}

\title{Semi-scaling cosmic strings}

\author{Vitaly Vanchurin}

\email{vanchurin@stanford.edu}

\affiliation{Department of Physics, Stanford University, Stanford, CA 94305}

\begin{abstract}

We develop a model of string dynamics with back-reaction from both scaling and non-scaling loops taken into account. The evolution of a string network is described by the distribution functions of coherence segments and kinks. We derive two non-linear equations which govern the evolution of the two distributions and solve them  analytically in the limit of late times. We also show that the correlation function is an exponential, and solve the dynamics for the corresponding spectrum of scaling loops.

\end{abstract}

\pacs{98.80.Cq	
	11.27.+d 
    }

\maketitle

\section{Introduction}

Topological defects, such as cosmic strings, could have formed at the end of the brane inflation \cite{Tye} or during the symmetry breaking phase transitions \cite{Kibble} in early universe. It is also known that the cosmic string could lead to a number of observable phenomena: gravitational lensing \cite{Vilenkin, book}, CMB non-Gaussianities \cite{FRSB,TNSYYS,Vanchurin2, Hindmarsh, RS}, gravitational waves \cite{DV} and ultra-high energy cosmic rays \cite{BV, BKV}. However before one can effectively look for the observational signatures from cosmic strings it is absolutely necessary to know the statistical properties of the strings at late cosmological times. Unfortunately this problem proved to be highly non-trivial and after thirty years of numerical \cite{BB,AS, BB91, VHS, RSB, VOV, VOV2, OV, MS} and analytical \cite{AV81, Kibble85, MartinsShellard, DPR, Vanchurin, ACK, CK}  studies the final word has not yet been said.

Nambu-Goto evolution of strings is usually described by decomposition of the position three vector $\bold{x}(\sigma, t)$ into right and left moving waves
\be
\bold{x}(\sigma, t) = \frac{\bold{a}(\sigma- t) +\bold{b}(\sigma + t) }{2}
\ee
with condition $|\bold{a}' |=|\bold{b}' |=1$,  where prime denotes a derivative with respect to $\sigma$. Without intersections the dynamics is completely linear in the flat space-time but some non-linearities already arise from cosmological stretching on the expanding backgrounds \cite{book}. Another source of non-linearities is due to the production of scaling and non-scaling loops \cite{Vanchurin3}. In order to analytically describe the complicated dynamics of cosmic strings all of the non-linear effects must be included in a self-consistent way. 

In the previous paper \cite{Vanchurin3} we made a first step to develop a self-consistent model by including the back-reaction from only {\it non-scaling} loops whose sizes do not scale with time. The main assumptions of the model were:
\\
1) non-scaling loops are predominantly produced at the scales of the initial correlation length $\sim l_{min}$. \\
2) large scale $\sim t$ inter-commutations do not significantly affect the statistical properties on smaller scales $\sim l_{min}$.\\
3) in the comoving coordinates $l_{min}$ remains constant, when all other length scales grow linearly with time.\\
4) the gravitational backreaction scale remains always smaller than the scale of initial conditions $\sim l_{min}$.\\
The first three assumptions are motivated by many numerical simulations \cite{RSB, VOV, VOV2, OV, MS} as well as by analytical results \cite{DPR, Vanchurin, Vanchurin2}, however,
the latter assumption might eventually break down. The effect of the gravitational backreaction on the string network is unimportant for comparisons of our model with Nambu-Goto numerical simulation \cite{RSB, VOV, VOV2, OV, MS}, but is very important phenomenologically and will be discussed extensively in a separate publication \cite{Vanchurin4}. 

The main mechanism responsible for production of non-scaling loops is described in details in the Appendix. To summarize, if one chooses at random two opposite moving wiggles (one left-moving and one right-moving) of invariant length $l_{min}$, then the probability of $\bold{a}'$ and $-\bold{b}'$ curves to intersect (or to form a cusp) is given by 
\be
p \sim (1-C(l_{min}))/2
\ee
 where 
 \be
 C(l) \equiv \langle \bold{a}'(0) \bold{a}'(l)\rangle
 \label{eq:correlation}
 \ee
 is a correlation function. In Ref. \cite{Vanchurin3} it was shown that the statistical properties of the network are uniquely determined by what we called the number of {\it directions}:
\be
N \equiv 1/p \sim 2/(1-C(l_{min})).
\ee

The non-scaling model predicts an appearance of two new length scales: the coherence length 
 \be
 \xi(t) = \frac{c}{N^2}t
  \label{eq:coherence}
 \ee
 and the cross-correlation length 
 \be
 \chi(t) = \frac{k }{N}t
 \ee 
 where $c \sim k \sim 1$, in addition to the well known inter-string distance and correlation length 
 \be
 d(t) \sim \zeta (t) \sim t
 \ee
 At the onset of evolution $N\sim 10$ and at late times $N$ grows logarithmically 
\be
N(t) \propto \log(t)
\ee
due to cosmological stretching and emission of small loops.

In this paper we make one step further and develop a semi-scaling model with back-reaction from both scaling and non-scaling loops. In a fully scaling network the total string length decays as
\be
\frac{d L(t)}{dt} = -\Gamma_{small} \frac{L(t)}{t}-\Gamma_{large} \frac{L(t)}{t}- \Gamma_{friction} \frac{L(t)}{t}.
\label{eq:overall_decay}
\ee
where
\be
\Gamma_{small} + \Gamma_{large} + \Gamma_{friction} =2
\ee
and the three terms describe the energy transfer into small loops, large loops and Hubble friction respectively. According to \cite{Vanchurin3}:
\be
\Gamma_{small} \sim \Gamma_{large} \sim 1
\label{eq:small_large}
\ee 
and  $\Gamma_{friction} \ll 1$. Thus, about half of the total energy goes into small loops and another half into large loops. Derivation of equations which describe both decay channels will be the main subject of this paper. In the second section we derive the evolution equation for the distribution of coherence segments and in the third section we derive the evolution equation for the angles between these segments.  The main results are summarized in the conclusion.

\section{Distribution of segments}

Consider the total number $n(\sigma,t)$ of coherence segments of length $\sigma$ at time $t$.  Without loss of generality we assume that non-scaling wiggles have unit size $l_{min} \sim 1$ and from the first assumption (see above) most of the non-scaling loops are of a unit size as well. Then there are three mechanisms which can make the number of segments $n(\sigma,t)$ change in time:

1) partial decay of a segment into unit size loops,

2) merger of a pair of segments into larger segments,

3) complete decay of segments into larger loops. \\

If we assume that on each unit time step only the leading unit size  (left- and right- moving) wiggle is removed from randomly chosen coherence segments, then from  (\ref{eq:coherence}), (\ref{eq:overall_decay}) and (\ref{eq:small_large}) the probability for a given segment of size $\sigma$ to shrink to a segment of size $\sim \sigma-1$ is given by
\be
P(\sigma-1 | \sigma) = \frac{L(t)/t}{\sum_{\sigma=1}^{\infty} n(\sigma,t)} =  \frac{\xi}{t} =  \frac{c}{N^2}.
\ee
where $\sum_{\sigma=1}^{\infty} n(\sigma,t)$ is the total number of coherence segments and $L(t)/t$ is the total number of unit wiggles to decay in unit size loops in unit time. 

We can now write down a finite difference equation with all three mechanisms taken into account:
\be
n(\sigma,t+1) - n(\sigma,t)= - n(\sigma,t) \frac{\xi}{t}  + n(\sigma+1,t) \frac{\xi}{t}  + \mu \sum_{x=1}^{\sigma-2} p(x,t) n(1,t) \frac{\xi}{t} p(\sigma-1-x,t) - \frac{n(\sigma,t)}{t}
\label{eq:finite_difference}
\ee
where 
\be
p(\sigma,t) =\frac{n(\sigma,t)}{\sum_{\sigma=1}^{\infty} n(\sigma,t)} = \frac{n(\sigma,t) \xi(t)}{L(t)}  \propto n(\sigma,t) t^3
\label{eq:segment_distribution}
\ee
is the probability for a random coherence segment to have length $\sigma$, and $\mu$ is the probability of the next-to-nearest coherence segments to point in the same direction which is required for a merger to occur. Because any two next-to-nearest segments had passed through almost the same set of opposite moving segments they are very likely to point in the same direction: $\mu \sim 1$. (For the dating toy model introduced in Ref. \cite{Vanchurin3} the number was calculated numerically: $\mu \sim 0.58$.) The first two terms on the LHS of (\ref{eq:finite_difference})  describe a partial decay of segments, the third term describes a merger of two segments pointing in the same direction and the last term describes the decay into large scaling loops. 

By plugging (\ref{eq:segment_distribution}) into (\ref{eq:finite_difference}) we get
\be
n(\sigma,t+1) - n(\sigma,t) \sim \frac{c}{N^2} \left ( n(\sigma+1,t) - n(\sigma,t) +  \frac{\mu \xi^2}{L(t)^2} \sum_{x=1}^{\sigma-2} n(x,t) n(1,t) n(\sigma-1-x,t) \right )- \frac{n(\sigma,t)}{t}
\ee
or in a differential form 
\be
\frac{d n(\sigma,t)}{dt}= - \frac{n(\sigma,t)}{t} + \frac{1}{N^2} \left (  C_1 \frac{dn(\sigma,t)}{d\sigma} + C_2 \frac{n(1,t)  \xi^2}{L(t)^2}  \int_1^{\sigma-2} n(x,t)  n(\sigma-1-x,t) dx \right )
\ee
where $C_1 \sim C_2 \sim 1$. We can use (\ref{eq:segment_distribution}) to write down an evolution equation for a probability distribution
\be
\frac{d p(\sigma,t)}{dt} = 2 \frac{p(\sigma,t)}{t}  + \frac{C_1}{N^2}  \frac{dp(\sigma,t)}{d\sigma} + \frac{C_2}{N^2}  p(1,t)  \int_1^{\sigma-2} p(x,t)  p(\sigma-1-x,t) dx.
\label{eq:evolution_equation}
\ee
Despite of its complexity the integro-differantial equation (\ref{eq:evolution_equation}) has a very simple solution in the limit $1 \ll \sigma \ll t$:
\be
p(\sigma,t) \propto \frac{\exp(-\frac{3 N^2 \sigma}{C_1 t})}{t} 										
\label{eq:evolution_solution}
\ee
which could be checked by direct substitution. In the expanding universe there would be an additional effect which could lead to a merger of two coherence length segments and would effectively lead to a faster growth of the coherence length or equivalently larger $C_1$. In fact this effect could be easily included in the non-linear term of (\ref{eq:evolution_equation}) with a proper redefinition of $C_2$, but the overall form of the solution (\ref{eq:evolution_solution}) would remain unaffected. 

Although the exponential distribution of coherence segments (\ref{eq:evolution_solution}) does not directly imply that the correlations function is necessarily an  exponential on all scales it certainly suggests that the latter is nearly linear on a wide range of scales between $l_{min}$ and $\xi$, where only the linear term in the expansion of the exponential gives us the dominant contribution. This feature is a direct prediction of our model and is mainly due to constant mergers of the next-to-nearest coherence segments described by a non-linear term in (\ref{eq:evolution_equation}).

From (\ref{eq:evolution_solution}) we can find the probability for a pair of unit size wiggles on a distance $\sigma$ to be belong to the same segment $\exp\left(-\frac{3 N^2 \sigma}{C_1 t}\right) \sim \exp\left(-\frac{\sigma}{\xi}\right)$ which can be used to estimate the probability for any two wiggles on a distance $\sigma \gg \xi$ to point in the same direction
\be
P(\sigma) \sim \exp\left(-\frac{\sigma}{\xi}\right) + \frac{\mu}{\xi^2}   \int_\xi^{\sigma-\xi} \exp\left(-\frac{x}{\xi}\right) \exp\left(-\frac{\sigma-\xi-x}{\xi}\right)  dx + ...
\ee
where the higher order terms have a similar structure. The resulting probability
\be
P(\sigma) \sim \exp\left(-\frac{\sigma}{\xi}\right) (1 + \mu \frac{\sigma}{\xi}  ... ) \sim  \exp\left(-\frac{(1-\mu) \sigma}{\xi} \right) ,
 \ee
but to get the correlation function one must sum over all $N$ directions which would still give us an exponential under assumption of Gaussianity
\be
C(\sigma) =  \exp\left(- \frac{\sigma}{a t}\right),
 \label{eq:exponential} 
 \ee
where
\be
\zeta \sim a t
\label{eq:correlation_length}
\ee
is the correlation length and $a \sim 1$. In what follows it will be convenient to consider a distribution of correlation segments $\tilde{\sigma}$ described by an exponential of (\ref{eq:exponential})
\be
\tilde{p}(\tilde{\sigma},t) \propto  \exp\left(-\frac{\tilde{\sigma}}{a t}\right ).
\label{eq:distribution_correlation}
\ee

\section{Distribution of kinks}

Nearby coherence segments are separated by sharp kinks  with angles $\theta > 1/\sqrt{N}$. To understand the distribution of $\theta$  we will derive an equation for the number density $n(\theta,t) d\theta$ in a sting network with total length in long strings $L(t)$. There are three mechanism which can lead to changes in $n(\theta,t)$: 

1) cosmological stretching, 

2) formation of loops,

3) large scale inter-commutations.\\
The first mechanism is described by 
\be
\frac{d\theta}{d t} \propto \alpha (2 \langle v^2 \rangle -1) \frac{\theta}{t},
\label{eq:stretching}
\ee
where the scale factor grows as $a(t) \propto t^\alpha$ and
\be
2 \langle v^2 \rangle -1 \sim - \frac{1}{N},
\label{eq:velocity}
\ee
according to \cite{Vanchurin3}. This is due to the fact that the opposite moving wiggles cannot point in all directions, or otherwise the small loops would have been formed. Note, that the corresponding power-law decaying  solution is valid only in the limit of large kinks $\theta \gg 1/\sqrt{N}$. If the cosmological stretching would be the only mechanism, then we would write $n(\theta+d\theta,t+dt) = n(\theta, t)$ or in a differential form
\be
\frac{d n(\theta,t)}{dt} = C_3 \frac{\alpha}{N} \frac{\theta}{t} \frac{d n(\theta,t)}{d\theta}. 
\ee
By definition of coherence segments \cite{Vanchurin3} , $n(\theta,t)$ must vanish for small kinks $\theta \ll 1/\sqrt{N}$ which can be imposed by an additional term
\be
\frac{d n(\theta,t)}{dt} = C_3 \frac{\alpha}{N} \frac{\theta}{t} \frac{d n(\theta,t)}{d\theta}  - C_4 \frac{1}{\theta^2 N} \frac{n(\theta,t)}{t} . 
\label{eq:only_stretching}
\ee
where $C_3$ and  $C_4$ are some constants of order one.

The second mechanism is due to unequal chances for different kinks to survive the evolution. Clearly the segments connected by small kinks have much larger probability to remain untouched when large kinks can easily form loops by interacting with similar opposite moving kinks. The probability for a random kink to have size $\theta$ is
\be
p(\theta,t) =  \frac{n(\theta,t) \xi(t)}{L(t)} \propto n(\theta,t) t^3.
\label{eq:distribution_kinks}
\ee
and the probability that in time $\xi$ it would meet an opposite moving kink of a similar size (on a logarithmic scale) is $\sim p(\theta,t) \theta$. If these two kinks also intersect on a unit sphere (which happens with probability $\propto \theta^2$) then we are almost guaranteed to create a loop (see Appendix). We can now generalize (\ref{eq:only_stretching}) to include the formation of intermediate loops
\be
\frac{d n(\theta,t)}{dt} = C_3 \frac{\alpha}{N} \frac{\theta}{t} \frac{d n(\theta,t)}{d\theta}  - C_4 \frac{1}{\theta^2 N} \frac{n(\theta,t)}{t} + C_5 \frac{n(\theta,t)^2}{L(t)} \theta^3- \frac{n(\theta,t)}{t}
\label{eq:stretching_loops}
\ee
where  $C_5 \sim 1$. The last terms comes from the assumption that the large loops can carry away random kinks without any preferences given to their angles $\theta$. 

Large scale inter-commutations constantly inject sharp kinks into the network. If the segments which undergo such intersections are completely uncorrelated, then the angles of kinks will be distributed as $\propto \sin(\theta)$. Clearly, it is easier to create kinks with angle $\sim \pi/2$ when two random segments intersect with frequency estimated as $\propto L(t)/t^2$. The overall evolution equation with all three mechanisms combined is given by
\be
\frac{d n(\theta,t)}{dt} = C_3 \frac{\alpha}{N} \frac{\theta}{t} \frac{d n(\theta,t)}{d\theta}  - C_4 \frac{1}{\theta^2 N} \frac{n(\theta,t)}{t} - C_5 \frac{n(\theta,t)^2}{L(t)} \theta^3 + C_6 \frac{L(t)}{t^2} \sin(\theta) - \frac{n(\theta,t)}{t}
\label{eq:stretching_loops_intercommutations}
\ee
where $C_6$ is yet another constant of order unity. 

We can substitute  (\ref{eq:distribution_kinks}) into (\ref{eq:stretching_loops_intercommutations}):
\be
\frac{d p(\theta,t)}{d\log(t)} = 2 p(\theta,t) + C_3  \frac{\alpha}{N} \theta \frac{d p(\theta,t)}{d\theta}  - C_4 \frac{1}{\theta^2 N} p(\theta,t) - C_5  \frac{N^2}{c} \theta^3 p(\theta,t)^2+ C_6  \frac{c}{N^2} \sin(\theta),
\label{eq:kinks_evolution}
\ee
and look for a stationary solution $p(\theta,t) = p(\theta)$:
\be
\frac{d p(\theta)}{d \theta} =  C_3 \frac{p(\theta)}{\theta^3}- C_4 N \frac{p(\theta)}{\theta}  + C_5 N^3 \theta^2 p(\theta)^2  - C_6 \frac{1}{ N} \frac{\sin(\theta)}{\theta}
\label{eq:kinks_stationary}
\ee
where $C_3$, $C_4$, $C_5$ and $C_6$  are now some other constants $\sim 1$. 

The first term on the RHS of (\ref{eq:kinks_stationary}) introduces a sharp exponential cut-off for the smallest angles $\ll 1/\sqrt{N}$ and can be neglected for large angles $\gg 1/\sqrt{N}$. Moreover, in the limit of late times when $N$ becomes significantly large the last term can also be dropped and we get a much simpler equation 
\be
\frac{d p(\theta)}{d \theta} =  - C_4 N \frac{p(\theta)}{\theta}  + C_5 N^3 \theta^2 p(\theta)^2.  
\label{eq:kinks_simplified}
\ee
For sufficiently large $N$ the decaying solution goes as:
\be
p(\theta) \propto \frac{1}{\theta^{3}}
\label{eq:kinks_solution}
\ee
which can be combined with  (\ref{eq:evolution_solution}) to obtain
\be
p(\theta,\sigma,t) \propto \frac{\exp(-\frac{3 N^2 \sigma}{C_1 t})}{t \theta^3} 		
\label{eq:solution}
\ee
in the limit $t\gg 1$, $1/\sqrt{N} \ll \theta \ll 1$ and $l_{min} \ll \sigma \ll t$,

For large scales $\sigma \sim t$ and large kinks $\theta \sim \pi/2$ the distributions $p(\sigma,t)$ and $p(\theta,t)$ do not give an accurate description of evolution since the loops consisting of many coherence segments could also form. Instead it is convenient to study the evolution of the correlation segments $\tilde{p}(\tilde{\sigma},t)$ and $\tilde{p}(\tilde{\theta}, t)$, one of which was already derived (\ref{eq:distribution_correlation}) without large-scale inter-commutations taken into account. In the limit of interest the large scale inter-commutations and formations of large loops are the main effects corresponding to the last two terms in (\ref{eq:kinks_stationary}). This equation can be rewritten for the angles between correlation segments 
\be
\frac{d \tilde{p}(\tilde{\theta})}{d \tilde{\theta}} =   C_3 N^3 \tilde{\theta}^2 \tilde{p}(\tilde{\theta})^2  - C_4 \frac{1}{ N} \frac{\sin(\tilde{\theta})}{\tilde{\theta}}.
\label{eq:large_kinks_stationary}
\ee
Clearly, the distribution function for large angles does not depend significantly on $\tilde{\theta}$ and thus, we can estimate the spectrum of large loops (defined in \cite{VOV2}) by the distribution of correlation segments with $\tilde{\theta} = \pi/2$:
\be
f(\tilde{\sigma},t) \propto \tilde{p}(\tilde{\sigma} ,\pi/2, t) \left(1- \left(\frac{\tilde{\sigma}}{b t}\right)^2\right)  
\label{eq:spectrum}
\ee 
where
\be
d(t) \sim b t.
\label{eq:correlation_length}
\ee
The second factor is due to the fact that large loops of size $\tilde{\sigma}$ could rejoin back to the network with probability proportional to their physical size $\sim \tilde{\sigma}^2$.

The large kinks are produced by intercommutations of nearby strings with about one kink per time $\sim d(t)/v$ per string length $\sim \zeta(t)$, where the average velocity $v$ is given by (\ref{eq:velocity}). We can assume that the probability to create a large kink on a correlation segment $\tilde{p}(\tilde{\sigma} ,\pi/2, t)$ is proportional to the area swept by the segment and to the distribution (\ref{eq:distribution_correlation}) generated by the dynamics on small scales. Then from (\ref{eq:spectrum}) and (\ref{eq:distribution_correlation}) the spectrum of loops is given by
\be
f(\tilde{\sigma},t)  \propto  \frac{\tilde{\sigma}}{t}\exp \left (- \frac{\tilde{\sigma} }{a t} -\left(\frac{\tilde{\sigma}}{b t}\right)^2\right ),
\label{eq:loops_spectrum}
\ee
with a scaling peak at around $\sigma \sim a t \sim b t$. 

\section{Conclusions}

The main point of the paper was to derive the evolution equations (\ref{eq:evolution_equation}) and  (\ref{eq:kinks_evolution}) with back-reaction from scaling and non-scaling loops taken into account. Remarkably the equations turned out to have very simple asymptotic solutions on a wide range of scales and sizes of kinks (\ref{eq:solution}). In addition, we obtained some preliminary results on the from of the correlation function (\ref{eq:exponential}) and on the spectrum of scaling loops (\ref{eq:loops_spectrum}). 

Throughout the paper we have explicitly assumed that the large scale inter-commutations and the small scale gravitational backreaction do not significantly affect the statistical properties on the scales of $l_{min}$. Although the assumption is very typical, it is not immediately clear why this should be the case in a non-linear system under investigation. Moreover it is already known that the opposite is not true and that the dynamics on the scales of $l_{min}$ affects the evolution on all scales \cite{Vanchurin3}. We are planning to study these and other related issues analytically and/or numerically in our future work \cite{Vanchurin4}.

\section*{Acknowledgments}

I am grateful to Alex Vilenkin for very helpful discussions.

\appendix
\section{Small loops}

To understand the formation of small loops we can consider the following decomposition of the tangent vectors
\bea
\bold{a}'(\sigma) =  \alpha(\sigma) \hat x +  \beta(\sigma) \hat y + \sqrt{1-\alpha(\sigma)^2 - \beta(\sigma)^2}\hat z \\
\bold{b}'(\sigma) = \gamma(\sigma) \hat x +  \delta(\sigma) \hat y -\sqrt{1-\gamma(\sigma)^2 - \delta(\sigma)^2}  \hat z 
\eea
where $\hat x$, $\hat y$ and $\hat z$ are the unit vectors. The above ansatz is quite general but nevertheless assumes that $\bold a'$ and $\bold b'$ have components along some direction (we call $z$ axis) with opposite signs. This assumption is well justified for small wiggles in the vicinity of cusps (at $\bold a'(\sigma)\sim -\bold b'(\sigma) \sim \hat z$). In this limit the four functions  ($\alpha, \beta, \gamma$ and $\delta$) are much smaller than one, and by emitting higher order terms we get
\bea
\bold{a}'(\sigma) =  \alpha(\sigma) \hat x +  \beta(\sigma) \hat y + \left (1-\frac{\alpha(\sigma)^2 + \beta(\sigma)^2}{2}\right) \hat z\\
\bold{b}'(\sigma) = \gamma(\sigma) \hat x + \delta(\sigma) \hat y  - \left (1-\frac{\gamma(\sigma)^2 + \delta(\sigma)^2}{2}\right) \hat z.
\label{eq:expansion}
\eea

A string self-intersects and a loop of size $l$ can form only when 
\be 
\frac{\bold a(\sigma-t)+\bold b(\sigma+t)}{2} = \frac{\bold a(\sigma-t+l)+\bold b(\sigma+t+l)}{2}.
\ee
It is convenient to define three functions
\bea
X(\sigma, t) &\equiv& \frac{\alpha(\sigma-t) +\gamma(\sigma+t)}{2}   
\label{eq:X}\\
Y(\sigma, t) &\equiv& \frac{\beta(\sigma-t) +\delta(\sigma+t)}{2} 
\label{eq:Y}\\
Z(\sigma, t) &\equiv&  \frac{-\alpha(\sigma-t)^2 - \beta(\sigma-t)^2+ \gamma(\sigma+t)^2 +\delta(\sigma+t)^2}{4}
\label{eq:Z}
\eea
and write down the system of three integral equations 
\bea
\int_{\sigma}^{\sigma+l} X(u, t)  du=0
\label{eq:system1}\\
\int_{\sigma}^{\sigma+l} Y(u, t)  du =0
\label{eq:system2}\\
\int_{\sigma}^{\sigma+l} Z(u, t)  du =0
\label{eq:system3}
\eea
with three unknown parameters $\sigma, l$ and $t$. It is clear that the solutions of the above system depends solely on the statistical properties of the three functions $X, Y$ and $Z$ which in turn are uniquely described by $\alpha, \beta, \gamma$ and $\delta$.

For the time being we can assume that the string is relatively straight on the scales $\sim \xi$ with only small wiggles with wavelength $\sim l_{min}$ and amplitude $~ P(l_{min}) l_{min}$. This means that all four functions $\alpha, \beta, \gamma$ and $\delta$ oscillate with wavelength $\sim l_{min}$ and amplitude $P(l_{min}) \ll 1$. If we ignore the cross correlations between opposite moving waves, then both $X$ and $Y$ must also be oscillating functions in each of the variables ($t$ or $l_{min}$) with similar wavelength and amplitude. However, the third function can be re-express as 
\be
Z(\sigma, t) =  \frac{\left (\gamma(\sigma+t)-\alpha(\sigma-t)\right) X(\sigma, t)  + \left(\delta(\sigma+t)- \beta(\sigma-t)\right) Y(\sigma,t)}{2}.
\label{eq:ZZ}
\ee
which shows that the amplitude of oscillations is much smaller $\sim P(l_{min})^2$, but the wavelength is still the same $\sim l_{min}$. In other words the variance on the scales $l_{min}$ is given by
\be
\sqrt{\langle X^2 \rangle - \langle X \rangle^2} \sim  \sqrt{\langle Y^2 \rangle - \langle Y \rangle^2} \sim P(l_{min})\;\text{and}
\;\sqrt{\langle Z^2 \rangle - \langle Z \rangle^2} \sim P(l_{min})^2.
\label{eq:variance}
\ee

According to (\ref{eq:ZZ}) we are guaranteed to have $Z(\sigma, t)=0$ in the instant moments when both function $X(\sigma, t)$ and $Y(\sigma,t)$ vanish. In fact, one can always rotate the axis such that $\alpha(0)=\beta(0)=\gamma(0) = \delta(0)=X(0, 0)=Y(0,0)=Z(0,0)=0$ which corresponds to a mini-cusp at $\sigma=0$ and $t=0$. Since we are not interested in the zeros of $X, Y$ and $Z$ but in the zeros of their integrals we should integrate our three functions over a finite range $l>0$ as in (\ref{eq:system1}),(\ref{eq:system2}) and (\ref{eq:system3}). In the limit of late times and small wiggles ($1/\xi \ll P(l_{min})$) we can expect the mean of oscillations to change very slowly
\be
\langle X \rangle \sim \langle Y \rangle \sim  \langle Z \rangle \sim  0.
\ee
which implies together with (\ref{eq:variance}) that the system of equations (\ref{eq:system1}),(\ref{eq:system2}) and (\ref{eq:system3}) must have about one solution per region $-l_{min} <  t  < l_{min}$, $-l_{min} < \sigma <l_{min}$ and $-l_{min} < l <l_{min}$ with $l\sim t \sim \sigma \sim l_{min}$. For $1/\xi \ll P(l_{min})$ at least one mini-cusps occurs per length $l_{min}$ and therefore the entire overlapping region of the curves $a'$ and $-b'$ is likely to decay into small loops. 

\end{document}